\pdfoutput=1
\PassOptionsToPackage{table,dvipsnames}{xcolor}
\documentclass[11pt]{article}

\usepackage[preprint]{acl}

\usepackage{times}
\usepackage{latexsym}
\usepackage{amssymb}
\usepackage{amsmath}

\usepackage[table,dvipsnames]{xcolor}

\usepackage{tcolorbox}
\usepackage{cleveref}
\usepackage{multirow}

\usepackage[T1]{fontenc}
\usepackage{subcaption} %
\usepackage[utf8]{inputenc}
\usepackage{booktabs} 
\usepackage{pifont}
\usepackage{microtype}

\usepackage{inconsolata}

\usepackage{graphicx}
\usepackage{hyperref}
\usepackage{algorithm}
\usepackage{algorithmic}
\usepackage{booktabs}

\definecolor{gray}{RGB}{236, 236, 236} 
\definecolor{lightgreen}{RGB}{32, 144, 140}  
\definecolor{lightorange}{RGB}{229, 231, 255} 
\definecolor{darkblue}{RGB}{15, 84, 205} 
\newcommand\method{\textsc{Rearank}}

\title{\method: Reasoning Re-ranking Agent via Reinforcement Learning }

\author{
   Le Zhang\textsuperscript{\rm 1,2}\thanks{equal contribution} \quad Bo Wang\textsuperscript{\rm 3$\ast$} \quad \textbf{Xipeng Qiu}\textsuperscript{\rm 3}\\ \textbf{Siva Reddy}\textsuperscript{\rm1,4,5}\quad \textbf{Aishwarya Agrawal}\textsuperscript{\rm 1,2,5} \\
  $^1$Mila - Quebec AI Institute \quad $^2$Université de Montréal  \quad $^3$Fudan University \\
    $^4$McGill University \quad
  $^5$Canada CIFAR AI Chair
}

\begin{document}
\maketitle
\begin{abstract}
We present \method, a large language model (LLM)-based listwise reasoning reranking agent. \textsc{Rearank} explicitly reasons before reranking, significantly improving both performance and interpretability. Leveraging reinforcement learning and data augmentation,  \textsc{Rearank} achieves substantial improvements over baseline models across popular information retrieval benchmarks, notably requiring only 179 annotated samples. Built on top of Qwen2.5-7B, our \method-7B demonstrates performance comparable to GPT-4 on both in-domain and out-of-domain benchmarks and even surpasses GPT-4 on reasoning-intensive BRIGHT benchmarks. These results underscore the effectiveness of our approach and highlight how reinforcement learning can enhance LLM reasoning capabilities in reranking. The
code is available \url{https://github.com/lezhang7/Rearank}.
\end{abstract}

\begin{figure}[t]
    \centering
    \includegraphics[width=\columnwidth]{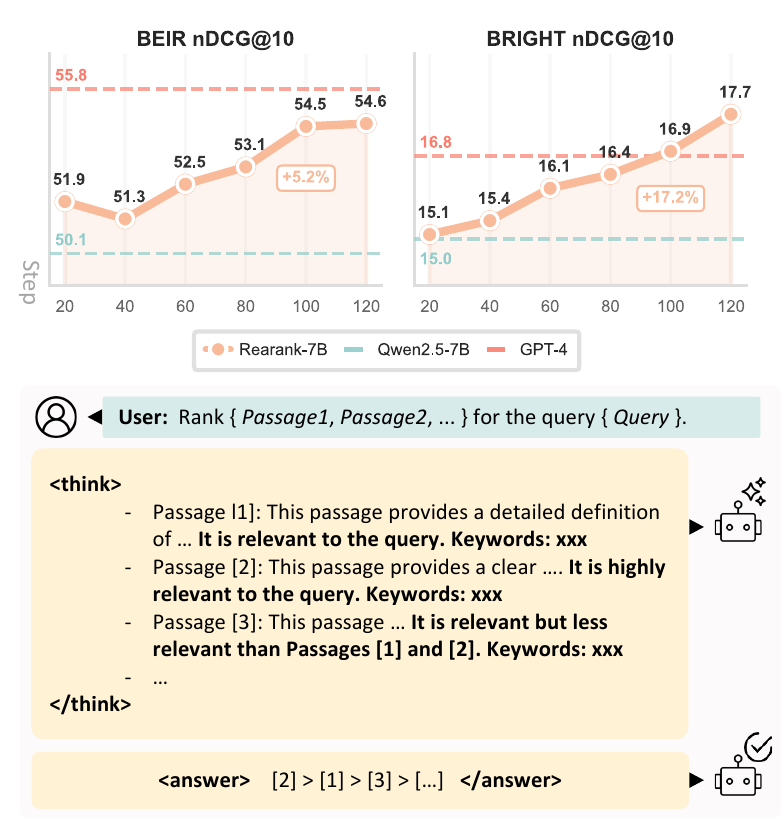}
    \caption{(Top) Average rerank results on popular benchmarks (over BM25 top 100), the performance improves with RL training; (Bottom) \textsc{Rearank} inference example. The agent provides the \textcolor{darkblue}{reasoning} and final ranking of all passages, unlike current agents \cite{rankgpt, pradeep2023rankzephyr} that only output the final answer.}
    \label{fig:overview}
    \vspace{-3mm}
\end{figure}

\section{Introduction}

Information retrieval (IR) is a core building block of intelligent systems, providing the foundation for accessing, organizing, and reasoning over information. Modern IR systems \cite{reimers-2019-sentence-bert, reimers-2020-multilingual-sentence-bert, wang2021bert, SPLADE} typically follow a two-stage approach: initial retrieval (e.g., fast lexical methods \cite{bm25}) to gather candidates, followed by reranking for fine-grained prioritization of relevant results. This two-stage process is particularly crucial for Retrieval-Augmented Generation systems \cite{lewis2020retrieval, borgeaud2022improving, zhang2023moqagpt}, where accurate retrieval and effective reranking of context passages significantly impact generated output quality.

Recent advances in large language models (LLMs) have shown strong promise for this reranking phase \cite{rankgpt}, particularly through their use as \textit{reranking agents} that rely on direct outputs rather than internal states (\textit{e.g. logits}) . This paradigm, exemplified by zero-shot prompting methods \cite{rankgpt, ma2023zero, zhuang2024setwise}, offers significant deployment flexibility, especially in model-as-a-service scenarios. However, effectively adapting LLMs specifically as reranking agents presents several key challenges: \textit{(i)} LLMs are not inherently optimized for ranking objectives, and crucially, current zero-shot methods do not learn from the interaction signals generated during the reranking process;
\textit{(ii)} Achieving competitive performance often necessitates supervised fine-tuning, a process severely constrained by the scarcity and high cost of acquiring high-quality labeled ranking data \cite{rankgpt};
\textit{(iii)} The decision-making processes within these models frequently lack transparent and interpretable reasoning, which limits explainability and fails to leverage test-time scaling properties of LLMs;
\textit{(iv)}  State-of-the-art reranking agents frequently depend on large, often proprietary models (e.g., GPT-4) or face significant challenges in local deployment.  This dependency on large models incurs substantial computational costs and significant inference latency (e.g., reranking 20 passages with reasoning process using DeepSeek-R1 \cite{guo2025deepseek} via API takes around 90-120 seconds).

In this work, we propose \textsc{Rearank}, the first reasoning \textit{listwise} reranking agent. \textsc{Rearank}'s optimization for listwise reranking with explicit reasoning is incentivized using reinforcement learning (RL). This approach effectively leverages the rich, order-based signals inherent in listwise reranking. To address the significant data scarcity challenge typically associated with training listwise models, we introduce a data augmentation pipeline capable of generating extensive listwise ranking sets from a remarkably small seed of only 179 annotated queries. At inference, \textsc{Rearank} generates explicit, interpretable reasoning for each ranking step, as shown in \cref{fig:overview}, enabling test-time scaling. Built upon the principle of injecting strong reasoning capabilities into a compact model, \textsc{Rearank} operates with low operational cost. The combination of its practical model size and listwise reranking strategy enhances inference efficiency by minimizing LLM calls, thereby facilitating local deployment.

Our experimental results demonstrate \textsc{Rearank}'s effectiveness: it significantly surpasses baseline models and achieves performance comparable to strong models like GPT-4 and the recent powerful reasoning model Qwen3-32B on standard and out-of-domain benchmarks. Notably, \textsc{Rearank} even surpasses strong GPT-4 performance on reasoning-intensive tasks, highlighting its advantages in combining reasoning with practical efficiency, as summarized in \cref{fig:overview}.

Our main contributions are threefold: \textit{(i) } We introduce \textsc{Rearank}, a novel reasoning-based reranking agent based on the listwise reranking strategy that effectively integrates explicit reasoning capabilities into the reranking process. We formularize the reranking problem in RL framework, and propose a data synthesis method requiring only 179 annotated queries, and a new reward model leveraging ranking information for training, enabling efficient RL training of \textsc{Rearank}; and \textit{(ii)} \textsc{Rearank} achieves significant performance improvements over strong baselines and matches or surpasses results from competitive models like GPT-4 and strong reasoning model Qwen3, particularly on reasoning-intensive tasks, while offering substantially improved inference efficiency due to its compact model size. \textit{(iii)} We provide a comprehensive analysis on reasoning transferability and examining the relationship between reasoning length and final ranking performance to better understand the role of reasoning in reranking.

\section{Related Work}

\paragraph{Large Reasoning Models}
Recent advancements in LLMs have yielded increasingly sophisticated reasoning capabilities, often emerging with scale \cite{wei2022emergent,kojima2022large}. Techniques like Chain-of-Thought prompting \cite{wei2022chain} and its variants \cite{kojima2022large,wang2022self} further enhance these skills by enabling explicit reasoning processes. Beyond prompting, training methods like RL are used to incentivize long0 CoT reasoning; models such as Deepseek-R1 \cite{guo2025deepseek}, OpenAI o1, and o3 leverage RL for enhanced reasoning, showing general task improvements. These developments enable LLM application in complex domains like math problems and planning. Our work applies these advanced reasoning capabilities to reranking.

\paragraph{LLMs for Re-ranking}
LLMs are increasingly being used for reranking, moving beyond traditional feature-based models \cite{zhang2024exploring, lee2024nv, ma2024fine, behnamghader2024llm2vec, sachan2022improving}. Approaches range from pointwise \cite{liang2022holistic}, pairwise \cite{qin2023large}, setwise \cite{zhuang2024setwise} to listwise methods \cite{ma2023zero,rankgpt, pradeep2023rankzephyr}. While few-shot/zero-shot prompting was explored, supervised fine-tuning on ranking data shows further gains \cite{pradeep2023rankzephyr,pradeep2023rankvicuna}. Building on this, our work proposes a novel RL approach without cold start for training a listwise LLM reranker, focusing on robust performance and reasoning capabilities, especially in out-of-domain and reasoning-intensive scenarios. 

\begin{figure}[thb]
    \centering
    \includegraphics[width=\columnwidth]{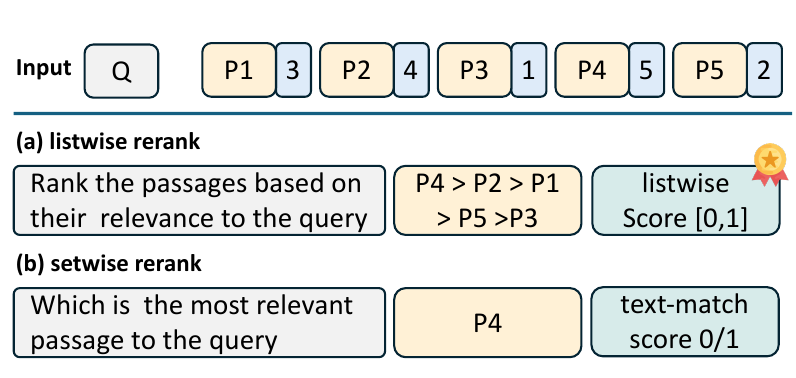}
\caption{\textbf{Listwise vs. Setwise Reranking.} Setwise reranking yields binary scores (0 or 1); listwise reranking offers richer, continuous scores between 0 and 1.}
    \label{fig:rerank example}
    \vspace{-3mm}
\end{figure}

A concurrent work \cite{rankr1} also trains an LLM as a \textit{setwise reranker} using RL. It simplifies the task to finding the \textit{single most relevant passage index}, relying on a sparse text-matching binary reward signal. This signal lacks the rich, order-based information present in listwise ranking, which consequently necessitates extensive training data. Furthermore, setwise inference is highly inefficient as it \textit{ranks only one passage at a time}, unlike our listwise method which \textit{reranks an entire passages simultaneously} as shown in \cref{fig:rerank example}. Therefore, we adopt the listwise reranking strategy.

\section{Method}


\subsection{Listwise Re-ranking Agent}

Given a query $q$ and an initial set of $n$ retrieved passages $P = (p_1, \ldots, p_n)$, the objective of reranking is to find the optimal permutation (ranking) of these passages. This can be formally expressed as maximizing a ranking quality score:
\begin{align}
\label{eq:reranking_objective}
\max_{\sigma \in K_n} r(\{p_{\sigma(1)}, p_{\sigma(2)}, \ldots, p_{\sigma(n)}\}),
\end{align}
where $K_n$ is the set of all possible permutations of the passages $P$, $\sigma$ represents a specific permutation (a ranking), $p_{\sigma(i)}$ is the passage located at rank $i$ in the ranking defined by $\sigma$, and $r$ is a scoring function that measures the quality of the entire ranked list.

Listwise reranking methods \cite{rankgpt,RankLlama, pradeep2023rankzephyr} reorder \textit{subsets of passages} using a sliding window. This is necessary due to the limited context length of LLMs, preventing simultaneous processing of all passage at a time. Given an initial ranking $\tau$ over passages $P$, an LLM-based permutation function $h$ (see \cref{fig:rerank example}) is applied to a window of size $w$ starting at index $i$ to determine the reordering within the final ranking $\sigma$ based on relevance to $q$:
\begin{equation}
    \begin{aligned}
&\{p_{\sigma(i)}, \ldots, p_{\sigma(i+w-1)}\}, \\ = & h(\{p_{\tau(i)}, \ldots, p_{\tau(i+w-1)}\}, q),
    \end{aligned}
\end{equation}

The final top-k list is constructed by iteratively applying $h$ using a sliding window that processes the whole passages list, often from the end towards the beginning. This window is typically shifted by $w/2$ steps to create overlap, resulting in approximately $O(2n/w)$ total LLM calls for $n$ passages and offering significant efficiency advantages. 
\label{sec: rerank}


\subsection{RL for Listwise Re-ranking}


A common mathematical framework for reinforcement learning is the Markov Decision Process (MDP), formally defined as a tuple $(S, A, T, r, \gamma)$. Here, $S$ represents the state space, $A$ is the action space, $T: \mathcal{S} \times \mathcal{A} \times \mathcal{S} \rightarrow [0, 1]$ denotes transition probabilities, $r: \mathcal{S} \times \mathcal{A} \rightarrow \mathbb{R}$ is the reward function, and $\gamma \in [0,1)$ is the discount factor.

In the context of passage reranking, we model the process as an MDP where the agent is our LLM policy $\pi_\theta$. The environment is defined by the query $q$ and an initial ranking $\tau$ over a set of passages $P = (p_1, \ldots, p_n)$. The state space $S$ is defined by the current ranking of passages and the query, specifically $\mathcal{S} = (\{p_{\tau(1)}, \ldots, p_{\tau(n)}\}, q)$. The action space $A$ corresponds to the set of possible permutation functions $h$ that the LLM can apply to the current state. The transition function $T$ models how actions lead to new states (rankings). The reward function $r$ quantifies the quality of a reranking action based on relevance metrics, providing feedback to the agent. 

We train $\pi_\theta$, an LLM fine-tuned to generate an output sequence $G$ consists of reasoning process and a new permutation $\sigma_\theta$ based on the input $x = (\{p_{\tau(1)}, \ldots, p_{\tau(n)}\}, q)$. The learning objective is to maximize the expected reward:
$$
\theta^* = \arg\max_\theta \mathbb{E}_{(q,P) \sim \mathcal{D}} [r(\sigma_\theta)],
$$
where $\mathcal{D}$ is the data distribution.

Inspired by DeepSeek-R1 \cite{guo2025deepseek}, we employ the simple Grouped Policy Optimization (GRPO) algorithm. Training involves sampling a group of output sequences $G=\{o_1, o_2, \dots, o_G\}$ for each input $x$ with the system prompt. A rule-based reward $r_i$ is computed for each output sequence $o_i$ and normalized within the group $G$ to yield advantages $\hat{A}_{i}$. Following DeepSeek-Math's \cite{shao2024deepseekmath} approach using current policy samples, the token-level objective is:
\begin{figure*}[!thb]
    \centering
    \includegraphics[width=0.8\linewidth]{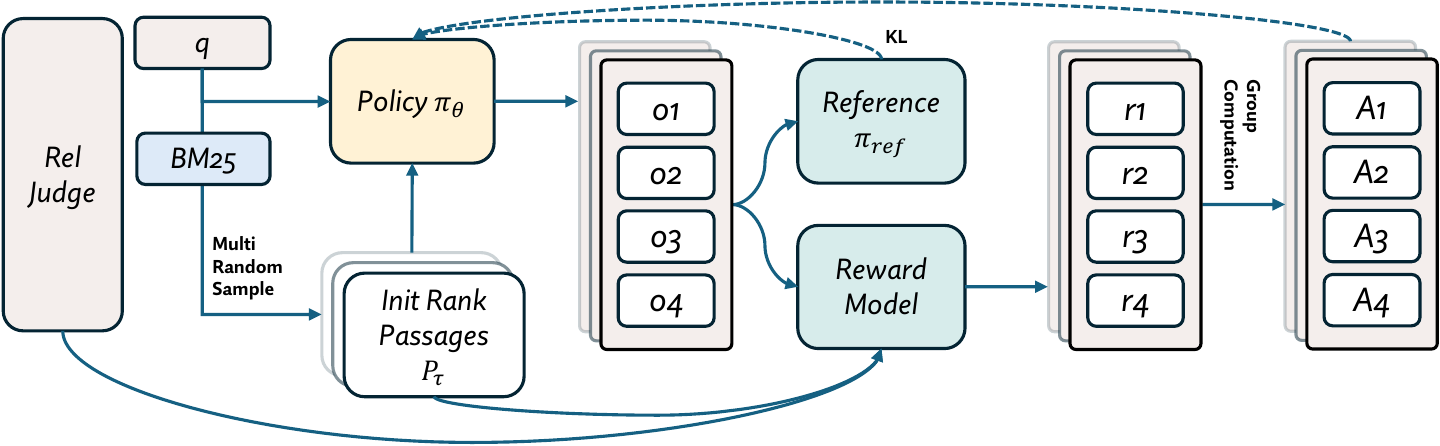}
\caption{Pipeline of the proposed GRPO-based RL framework for listwise passage reranking. Training utilizes data generated by sampling multiple passage sets per query and evaluating them with consistent relevance judgments. }
    \label{fig:grpo}
    \vspace{-3mm}
\end{figure*}
\begin{equation}
    \mathcal{J}_{\mathrm{GRPO}}(\theta)=-\frac{1}{|G|} \sum_{i=1}^{|G|} \frac{1}{\left|o_i\right|} \sum_{t=1}^{\left|o_i\right|} l_{i, t},
\end{equation}
where
\begin{equation}
    \begin{aligned}
        l_{i, t} &= \frac{\pi_\theta\left(o_{i, t} \mid x, o_{i,<t}\right)}{\left[\pi_\theta\left(o_{i, t} \mid x, o_{i,<t}\right)\right]_{\mathrm{nograd}}} \hat{A}_{i, t} \\
        &-\beta \mathbb{D}_{\mathrm{KL}}\left[\pi_\theta(\cdot \mid x, o_{i,<t}) \| \pi_{\mathrm{ref}}(\cdot \mid x, o_{i,<t})\right].
    \end{aligned}
\end{equation}

Here, $|G|$ is the number of output sequences sampled from current policy $\pi_\theta$ given input $x$, $|o_i|$ is sequence length, and $l_{i,t}$ is the per-token loss. $\pi_\theta(o_{i,t} \mid x, o_{i,<t})$ is the policy probability for token $o_{i,t}$ given $x$ and preceding tokens $o_{i,<t}$, with $[\cdot]_{\mathrm{nograd}}$ denoting gradient detachment. The token-wise advantage $\hat{A}_{i,t}$ is derived from outcome supervision,  calculated as the Z-score of the instance reward $r_i$ relative to the batch rewards $\mathbf{r}$: $\hat{A}_{i, t}=\frac{r_i-\operatorname{mean}(\mathbf{r})}{\operatorname{std}(\mathbf{r})}$. The KL penalty, using a fixed reference policy $\pi_{\mathrm{ref}}$ and coefficient $\beta$, encourages stability. The algorithm pipeline is shown in \cref{fig:grpo}.


\subsection{Reward Design}
The reward signal guides the RL agent by evaluating the quality of its generated output sequence. The output sequence $G_i$ includes structured components for reasoning ($\texttt{<think>...<|think|>}$) and ranking ($\texttt{<answer>...<|answer|>}$), as illustrated in \cref{fig:overview}. The total reward $r$ is a composite signal designed to encourage both high ranking performance and adherence to the desired output format.

The primary reward signal is based on the rich, order-based information inherent in listwise reranking, measured by Normalized Discounted Cumulative Gain (NDCG). Considering LLM context limits, we use NDCG@10 to evaluate the ranking of top-10 passages. For a generated permutation $\{p_{\sigma(1)}, p_{\sigma(2)}, \ldots, p_{\sigma(n)}\}$ for $q$, its score is $r_\text{rerank} = \text{NDCG}@10 = \frac{\text{DCG}@10}{\text{IDCG}@10}$, where $\text{DCG}@10 = \sum_{i=1}^{10} \frac{rel_i}{\log_2(i+1)}$ and $\text{IDCG}@10 = \sum_{i=1}^{10} \frac{rel_{i}^{\text{ideal}}}{\log_2(i+1)}$. Here, each query has an associated relevance judgements including a set of passages with scores annotated by human expert, $rel_i$ is the relevance score from the annotated relevance judgments for $p_{\sigma(i)}$ in the generated permutation, and $rel_{i}^{\text{ideal}}$ is the ideal relevance score from the relevance judgments for the passage that would be at rank $i$ in the ideal ranking for that query. Since the maximum possible NDCG@10 can vary depending on whether the randomly sampled passage set $P$ contains relevant documents, we define the ranking reward $r_\text{rank}$ as a \textbf{relative improvement score}. This approach uses min-max normalization to normalize for the differences in scales of reward scores between two different candidate sets and reduces reward variance \cite{greensmith2004variance, schulman2015high}: 

\begin{equation} 
r_\text{rank} = \frac{r_\text{rerank}-r_\text{init}}{r^*-r_\text{init}},
\end{equation}where $r^*=\max_{\sigma \in K_n} r(\{p_{\sigma(1)}, p_{\sigma(2)}, \ldots, p_{\sigma(n)}\})$ is the best achievable NDCG@10 for that specific passages set $P$; $r_\text{init} = r(\{p_{\tau(1)}, p_{\tau(2)}, \ldots, p_{\tau(n)}\})$ is the NDCG@10 of the initial ranking $\tau$ over $P$.

We also incorporate format rewards to encourage the desired output structure. A reward $r_\text{format1}=1$ is given if both $\texttt{<think>}$ and $\texttt{<answer>}$ tags are present in the output sequence. A reward $r_\text{format2}=1$ is given if the content within the $\texttt{<answer>}$ tags follows the specified ranking list format (e.g., $\texttt{[3] > [1] >[2]}$).

The final reward $r$ for a generated sequence is a weighted sum of these components:
\begin{equation}
    r = 0.8 \cdot r_\text{rank} + 0.1 \cdot r_\text{format1} + 0.1 \cdot r_\text{format2}
\end{equation}

\subsection{Initial State Expansion}

Training effective listwise rerankers faces a primary challenge in the scarcity of high-quality training data, defined as query $q$ paired with relevance judgments. To address this, we introduce a multi-sampling data augmentation method. Utilizing a small dataset of 179 queries from MSMARCO-V2 with fine-grained relevance judgments (0-3), we randomly sample multiple diverse sets of candidate passages $P$ for each query $q$ from its BM25 top 100 retrieval results. The core idea is to evaluate new ranking $\{p_{\sigma(1)}, \ldots, p_{\sigma(n)}\}$ produced from these varied initial ranking $\tau$ over diverse $P$ using the same set of relevance judgments. This multi-sampling approach shares conceptual similarities with negative sampling in contrastive learning \cite{xiong2020approximate, Zhang_2024_CVPR, zhang2024assessing, behnamghader2024llm2vec, lee2024nv}. By sampling varied initial passage sets and intial ranking, we effectively generate diverse ranking scenarios for a given query, allowing the model to learn robustly from a wider range of non-ideal inputs. This method generates rich listwise ranking data from limited annotations, enabling the model to learn robustly from diverse initial conditions and significantly reducing the need for large-scale, fully annotated query sets.


\section{Experiments}

\subsection{Experimental Setup}

\begin{table*}[thbp] 
  \centering

   \resizebox{\linewidth}{!}{%
  \begin{tabular}{llllllllllll} 
    \toprule 
    \multirow{2}{*}{\textbf{Model}} & \multirow{2}{*}{\textbf{\#Train}} & \multicolumn{2}{c}{\textbf{In-Domain}} & \multicolumn{8}{c}{\textbf{Out-of-Domain}} \\ 
    \cmidrule(lr){3-4} \cmidrule(lr){5-12} 
    &  & DL19 & DL20 & Covid & NFCorpus & DBPedia & SciFact & Signal & News & Robust04 & BEIR (Avg) \\
    \midrule 
   \rowcolor{gray} BM25\dag & - & 50.58 & 47.96 & 59.47 & 30.75 & 31.80 & 67.89 & 33.05 & 39.52 & 40.70 & 43.31 \\ 
    $\text{RankQwen}_\text{2.5}\text{-7B}$ & zero-shot & 68.25 & 62.73 & 77.74 & 37.40 & 39.83 & 70.83 & 31.73 & 43.24 & 49.95 & 50.10 \\
    $\text{RankGPT}_\text{3.5}$ & zero-shot & 65.80 & 62.91 & 76.67 & 35.62 & 44.47 & 70.43 & 32.10 & 48.85 & 50.62 & 51.25 \\
    $\text{RankGPT}_\text{4} \ddag$  & zero-shot & 75.59 & 70.56 & 85.51 & 38.47 & 47.12 & 74.95 & 34.40 & 52.89 & 57.55 & 55.84 \\
   RankZephyr-7B & 105K SFT & 73.90 & 70.60 & 83.54	& 38.38	&44.34	&75.18 &	31.44&	52.35&	54.19	&54.20 \\ \midrule
   \multicolumn{12}{c}{\textit{\textbf{Reasoning Language Model}}} \\
    $\text{RankQwen}_\text{3}\text{-32B}$ & zero-shot & 73.13 & 70.00& 83.86 & 36.28 & 45.44 & 71.78 & 32.06 & 51.73& 	57.27& 	54.06 \\
    $\text{RankQwen}_\text{3}\text{-235B}$ & zero-shot & 71.94& 69.37& 83.68 & 35.64& 41.33 & 63.30 & 32.53 & 50.79 & 58.16 & 52.20 \\
    Rank-R1-7B & 72k RL & 72.70 & 68.50 & 83.12 &	35.97&	43.43&	74.47&	32.16&	48.43&	55.17&	53.25 \\
   \rowcolor{lightorange!40}    \method-7B & \textbf{179} RL & 74.16 & 70.00 & 81.28 & 35.20 & 45.23 & 75.02 & 36.00 & 51.88 & 57.49 & 54.59 \\
   \rowcolor{lightorange!40} \quad $\Delta$  over baseline* &  &+5.91&	+7.27&	+3.54&	-2.20&	+5.40&	+4.19&	+4.27&	+8.64&	+7.54&	+4.49  \\
    \bottomrule 
  \end{tabular}%

  }
\caption{\textbf{Reranking Agent Results (nDCG@10) on TREC-DL and BEIR Benchmarks.} $\dag$ denotes initial BM25 retrieval performance. $\ddag \text{RankGPT}_\text{4}$ reranks top 30 passages from  $\text{RankGPT}_\text{3.5}$. All other models rerank on the BM25 top 100 passages. Training data size indicates the number of annotated queries used. *Baseline is $\text{RankQwen}_\text{2.5}\text{-7B}$}
  \label{tab:result1}
\vspace{-3mm}
\end{table*}

\paragraph{Training Details} Training data instances are generated by randomly sampling 20 candidate passages per query, repeated 50 times for each of the 179 queries. Samples without any passages marked as relevant (with a score = 0) in the relevance judgments or the initial nDCG@10 < 0.1 are filtered out, resulting in 12k training instances. \textit{We choose Qwen2.5-7B-Instruct as our baseline model.} Training is conducted using the VeRL \cite{verl} framework with a batch size of 128 and 32 rollouts per step. We trained the model directly via RL, without an initial SFT phase. The training runs for over 160 steps across 8 H100 GPUs.

\paragraph{Sliding window reranking} Our \textsc{Rearank} employs a sliding window to get top-k passages as introduced in \cref{sec: rerank}. Following common practice \cite{rankgpt, rankr1, zhang2024exploring}, we retrieve the top 100 passages using BM25 with plain query ($n=100$). We set the window size to $k=20$, resulting in 10 LLM calls per query ($2 \times 100 / 20 = 10$) to get top-10 passages.

\paragraph{Baselines} To evaluate the effectiveness of \textsc{Rearank}, we compare it against several strong baselines representing different reranking paradigms. Our zero-shot baselines include powerful large language models: Qwen2.5-7B \cite{yang2024qwen2} and GPT-4 \cite{gpt4}. We also incorporate the state-of-the-art open-source reasoning language model Qwen3-32B \cite{yang2025qwen3} and Qwen3-235B-A22B, which claim to surpass 671B Deepseek-R1. We adapt these models using the same sliding window strategy and prompt as \textsc{Rearank}, naming them RankQwen and $\text{RankGPT}$, respectively. As a supervised fine-tuning (SFT) baseline, we include RankZephyr \cite{pradeep2023rankzephyr}, which is distilled on 105k synthetic ranking data sourced from RankGPT. Furthermore, to compare different RL training strategies, we include Rank-R1 \cite{zhuang2024setwise}, a concurrent LLM reranker using the same base model (Qwen2.5-7B) but adopting a setwise reranking strategy. This provides valuable insights into the differences between the listwise (used by \textsc{Rearank}) and setwise RL approaches for reranking. The prompts are shown at app. \ref{sec:prompt}.

\paragraph{Benchmarks} To thoroughly evaluate our reasoning-enhanced reranker's performance and generalization, we select three distinct benchmark suites. We evaluate on the in-domain TREC-DL19 \cite{dl19} and DL20 \cite{dl20} datasets, both derived from MS-MARCO-V1. For assessing out-of-domain (OOD) generalization, we use BEIR \cite{thakur2021beir}, a diverse collection from sources outside of MS-MARCO. Crucially, real-world information retrieval tasks often demand capabilities beyond simple keyword matching or semantic similarity, requiring deeper reasoning to understand complex relationships or logic within the content and the query. To systematically evaluate our reasoning-enhanced \method, we utilize the BRIGHT \cite{su2024bright} benchmark, which is specifically designed to test reasoning abilities in retrieval contexts. For all evaluations, we report the nDCG@10 as the evaluation metric.

\subsection{In-domain \& OOD Retrieval Results}

As shown in \cref{tab:result1}, GPT-4 achieves the best performance across benchmarks, due to its superior text understanding. The Qwen2.5-7B also performs strongly, surpassing the legacy GPT-3.5 on both sets. Our \method, based on Qwen2.5-7B trained via our RL approach with reasoning ability, demonstrates performance closely comparable to GPT-4. This suggests the significant potential of advanced reasoning capabilities learned via RL for reranking.

\textsc{Rearank} achieves considerable improvements over the RankQwen2.5-7B baseline: a significant \textit{6.5\% improvement in nDCG@10} on the in-domain benchmarks and a notable \textit{4.5\% improvement in nDCG@10} on the OOD benchmark. These substantial gains are particularly impressive given that they are achieved using only 179 annotated queries (used for RL training). 

Comparing against SFT RankZephyr-7B, \textsc{Rearank}-7B demonstrates comparable performance on TREC-DL while achieving superior results in BEIR. This finding suggests that while SFT can be effective for in-domain data, our RL approach may offer enhanced robustness and better generalization capabilities when applied to out-of-domain tasks.

Evaluating against powerful reasoning language models, including the large Qwen3 and the concurrent Setwise Rank-R1, reveals significant strengths of \method. Despite being trained on a dataset of only 179 queries (0.2\% of Rank-R1's reported data), \method-7B surpasses Setwise Rank-R1 across benchmarks. Furthermore, \method-7B's performance exceeds both Qwen3-32B and Qwen3-235B. \method-7B achieves superior reranking performance with less training data and a more compact size than state-of-the-art methods. Interestingly, Qwen3-32B surpasses Qwen3-235B. Our investigation, consistent with \cite{marjanovic2025deepseek}, suggests Qwen3-235B's excessive self-reflection (marked by "\texttt{wait}") leads to confusion and degraded performance.

 \begin{table*}[!htbp] 
  \centering
  \resizebox{\linewidth}{!}{
  \begin{tabular}{lllllllllllllll} 
    \toprule 
    \multirow{2}{*}{\textbf{Model}}& \multirow{2}{*}{\textbf{\#Train}} & \multicolumn{7}{c}{\textbf{StackExchange}} & \multicolumn{2}{c}{\textbf{Coding}} & \multicolumn{3}{c}{\textbf{Theorem-based}} &  \multirow{2}{*}{\textbf{Avg.}}\\ 
    \cmidrule(lr){3-9} \cmidrule(lr){10-11} \cmidrule(lr){12-14} 
    & & Bio. & Earth. & Econ. & Psy. & Rob. & Stack. & Sus. & Pony & LC. & AoPS & TheoT. & TheoQ. & \\
    \midrule 
    \rowcolor{gray}BM25\dag{} & - & 18.2 & 27.9 & 16.4 & 13.4 & 10.9 & 16.3 & 16.1 & 4.3 & 24.7 & 6.5 & 2.1 & 7.3 & 13.7 \\ 
    $\text{RankQwen}_\text{2.5}\text{-7B}$ & zero-shot & 22.7 & 25.8 & 14.6 & 18.7 & 14.2 & 11.7 & 21.4 & 5.3 & 23.9 & 6.0 & 7.4 & 7.9 & 15.0 \\
    $\text{RankGPT}_\text{4}$ & zero-shot & 33.8 & 34.2 & 16.7 & 27.0 & 22.3 & 27.7 & 11.1 & 15.6 & 3.4 & 1.2 & 8.6 & 0.2 & 16.8 \\ 
    RankZephyr-7B & 105K SFT& 21.9 & 23.7 & 14.4 & 10.3 & 7.6 & 13.7 & 16.6 & 6.5 & 24.7 & 6.8 & 2.0 & 7.3 & 13.0  \\ \midrule
   \multicolumn{15}{c}{\textit{\textbf{Reasoning Language Model}}}  \\
   $\text{RankQwen}_\text{3}\text{-32B}$ & zero-shot & 24.9&	29.4&	20.9&	25.7&	18.3&	16.0&	23.2&	7.6&	27.6&	7.8&	8.9&	8.4&18.2\\
    $\text{RankQwen}_\text{3}\text{-235B}$ & zero-shot & 26.4 & 26.7&	22.1&	26.3	&18.8&	17.0&	24.9&	8.2	&27.2&	7.7&	11.7&	8.6&	18.8  \\
    Rank-R1-7B & 72k RL & 26.0 & 28.5 & 17.2 & 24.2 & 19.1 & 10.4 & 24.2 & 4.3 & 19.8 & 4.3 & 10.9 & 8.3 & 16.4 \\
    \rowcolor{lightorange!40}\textsc{Rearank}-7B & 179 RL & 23.4 & 27.4 & 18.5 & 24.2 & 17.4 & 16.3 & 25.1 & 8.0 & 27.0 & 7.4 & 9.5 & 7.9 & 17.7 \\
      \rowcolor{lightorange!40} \quad $\Delta$  over baseline* & & +0.8& 	+1.7& 	+4.0& 	+5.5& 	+3.3& 	+4.6& 	+3.7& 	+2.7& 	+3.0& 	+1.4& 	+2.1& 	0.0& 	+2.7\\
    \bottomrule 
  \end{tabular}%
  }
\caption{\textbf{Reranking Agent Results (nDCG@10) on BRIGHT .} \dag{} represent initial retrieval; all other models show reranking performance on the top 100 BM25 results. *Baseline is $\text{RankQwen}_\text{2.5}\text{-7B}$}
  \label{tab:result2}
  \vspace{-3mm}
  
  \end{table*}

\subsection{Reasoning-intensive Retrieval Results}
Table \ref{tab:result2} presents performance results on the reasoning-intensive BRIGHT benchmark. Notably, \textsc{Rearank}-7B even outperforms the powerful GPT-4 model on this benchmark, highlighting its strong reasoning capabilities developed through RL training. The SFT method, RankZephyr-7B, performs considerably poorer on BRIGHT, falling below the lexical-based BM25 baseline. This reinforces our observation that SFT generalizes poorly to out-of-domain and reasoning-intensive scenarios.

Comparing with the concurrent Setwise Rank-R1, our listwise \textsc{Rearank}-7B achieves better performance on the BRIGHT benchmark also. This superior performance, particularly against another RL-trained model, underscores the effectiveness of our approach. Furthermore, while \method-7B is smaller and does not undergo general reasoning training (\textit{e.g. math, coding, agent}) like Qwen3, its performance remains closely comparable. This near parity, despite inherent disadvantages, further confirms our method's effectiveness.

These strong results on a reasoning-intensive task against competitive LLM baselines (GPT-4, Rank-R1) highlight the efficacy of \textsc{Rearank}'s design. Its RL framework, trained on a limited high-quality dataset via our synthesis pipeline, enables learning complex reranking strategies. The listwise nature provides a richer signal for robust ranking compared to setwise methods, while also offering inference efficiency (fewer LLM calls).

\begin{table}[!t]
  \centering
   \resizebox{\linewidth}{!}{%
  \begin{tabular}{lccc}
    \toprule
    \textbf{Model Variant} & \textbf{TREC-DL} & \textbf{BEIR} &  \textbf{BRIGHT} \\ \midrule
    Qwen2.5-7B (baseline) & 65.5 & 50.1 & 15.0  \\ 
    \quad +Reasoning Prompt & 65.9 & 51.4 & 15.4 \\ \midrule
    wo/ Filter nDCG@10 < 0.1 & 71.3 & 53.6 & 16.9   \\
    w/ $r_\text{rank} = S_\text{rerank}$ & 70.9 & 53.2 & 16.7 \\
    w/ $r_\text{rank} = S_\text{rerank} - S_\text{init}$ & 71.5 & 54.0 & 17.2 \\

    w/ direct SFT & 66.7 & 50.7 & 14.7 \\

    \midrule
    \method-7B (full model) & 72.0 & 54.6 & 17.7 \\
    \bottomrule
  \end{tabular}
  }
  \caption{\textbf{Ablation of components of the approach.}}
  \label{tab:ablation}
  \vspace{-5mm}
\end{table}

\subsection{Ablation Studies}

To quantify the effectiveness of individual components within \method, we conducted ablation studies, summarized in \cref{tab:ablation}. Our first investigation examined the effect of applying the \textsc{Rearank}'s reasoning prompt directly to Qwen2.5-7B. While Qwen2.5-7B shows marginal improvement with the reasoning prompt over its zero-shot baseline, it significantly underperforms \method. This suggests that prompting alone is insufficient for eliciting robust reasoning in this context.

We then investigate the impact of data filtering. The "wo/ Filter nDCG@10 < 0.1" variant removes the filter on low-quality candidate sets (best possible nDCG@10 < 0.1)—many lacking relevant passages and yielding zero reward—which degrades performance. This underscores the importance of curating high-quality training instances.

Next, we explored reward function design. Using raw NDCG@10 ("w/ $r_\text{rank} = S_\text{rerank}$") results in lower performance due to high variance. Subtracting the initial score ("w/ $r_\text{rank} = S_\text{rerank} - S_\text{init}$") improved stability but still underperforms \method. This could be due to small learning signals as a result of small reward value scale. Our full model's normalized reward function provided more effective guidance, yielding the best results.

Finally, to isolate the benefits of our RL training approach from those gained solely from training on high-quality data, we also trained a SFT baseline ("w/ direct SFT"). This model was trained on 12k instances representing the best possible rankings derived from our multi-sampled candidate sets based on relevance judgments. Trained on this small dataset, the SFT baseline showed marginal improvements on in-domain and OOD tasks compared to the base Qwen2.5-7B model, but importantly, it negatively impacted the model's reasoning ability on the BRIGHT benchmark. This underscores the necessity of the RL approach for effectively leveraging small, high-quality data to train a robust and reasoning-capable reranking agent.



\begin{figure}[!thb]
    \centering
    \includegraphics[width=0.9\columnwidth]{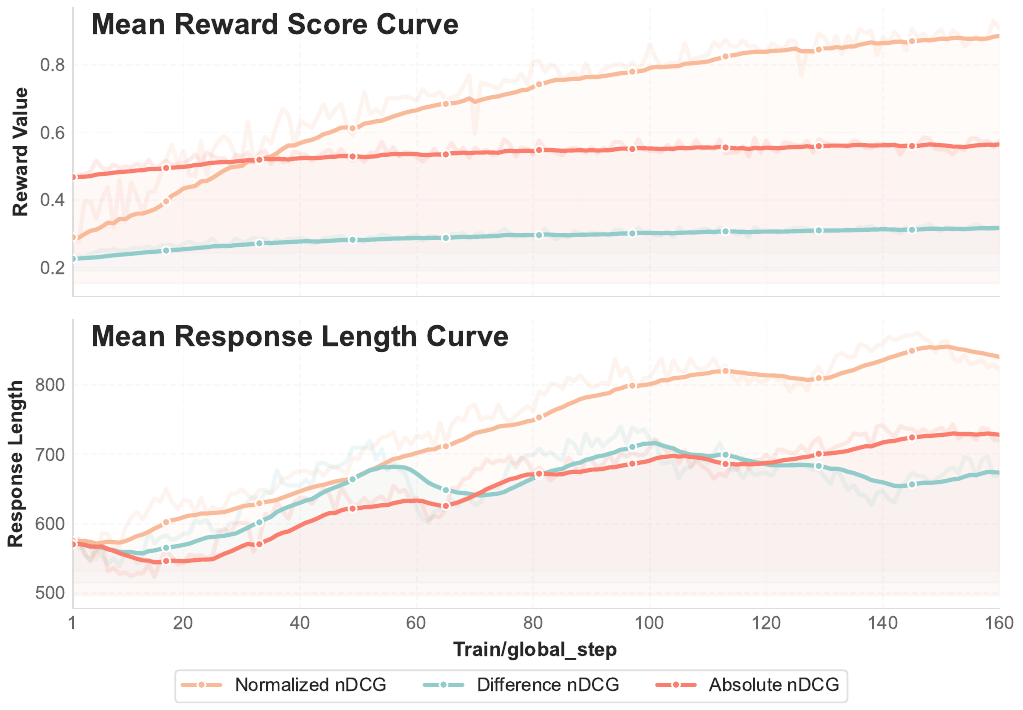}
\caption{(Top) Reward evolving curve (Bottom) Response length curve. }
    \label{fig:curve}
    \vspace{-3mm}
\end{figure}

\section{Analysis}

\paragraph{Analysis of Reward Functions}
We investigate the impact of different reward functions on \textsc{Rearank}'s training. We compare three nDCG@10 formulations: \textbf{Normalized nDCG} (our method: scaled by ideal nDCG), \textbf{Absolute nDCG} ($S_\text{rerank}$), and \textbf{Difference nDCG} ($r_\text{rank} = S_\text{rerank} - S_\text{init}$, improvement over initial).

The top panel of Figure \ref{fig:curve} shows reward curves. Normalized nDCG consistently grows, reaching $\sim 0.8$ by 100 steps, signifying that the achieved ranking quality is approximately 80\% of the ideal quality for the candidate set. Absolute nDCG saturates early ($\sim 50$ steps). Difference nDCG shows limited progress, plateauing at $\sim 0.15$ average improvement. This suggests Normalized nDCG provides a more effective signal for learning towards optimal ranking quality. The bottom panel shows response length trends. Normalized nDCG yields generally longer responses ($\sim 850$ tokens). Other functions saturate earlier ($\sim 700$ tokens), suggesting that the stronger signal from Normalized nDCG encourages more detailed responses and reasoning. 

\begin{figure*}[!thb]
    \centering
    \includegraphics[width=\linewidth]{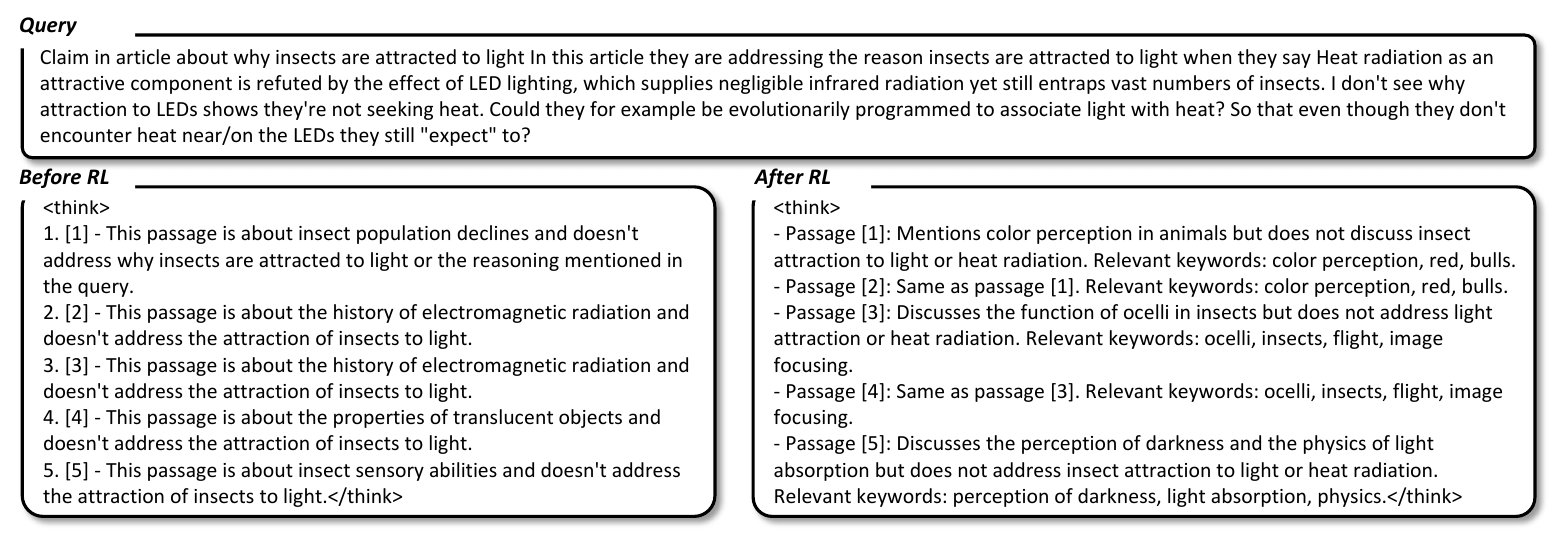}
\caption{Reasoning patterns: Before- vs. After-RL training under identical prompt and query.}
    \label{fig:reason_patter}
    \vspace{-3mm}
    \end{figure*}

\paragraph{Is reasoning helpful?}

The reasoning ability of \method\ is activated by the system prompt used in RL training (see app.\ref{sec:prompt}).  While Qwen3-32B also provides a switch to enable or disable the "thinking" mode, our observations, detailed in \cref{tab:reason}, indicate that enabling reasoning yields only marginal improvements in its reranking performance.  Specifically, Qwen3-32B shows slight gains on TREC-DL and BRIGHT (+0.6 and +0.4, respectively) and a minor decrease on BEIR (-0.2), suggesting \textit{its high performance stems primarily from strong pre-training, rather than reasoning capacity.}

In contrast, reasoning is core to our specialized reranking agent, \method, and significantly enhances its performance.  As the table illustrates, \method-7B without reasoning (use zero-shot prompt), already outperforms the Qwen2.5-7B baseline, demonstrating an inherent improvement in reranking capacity.  However, incorporating reasoning leads to considerable gains, underscoring the importance of reasoning learned via RL.

\begin{table}[!t]
  \centering
   \resizebox{\linewidth}{!}{%
  \begin{tabular}{lcccc}
    \toprule
    \textbf{Model} & \textbf{Reasoning} &\textbf{TREC-DL} & \textbf{BEIR} &  \textbf{BRIGHT} \\ \midrule
    Qwen3-32B & \ding{51} & 71.6&	54.1&	18.2 \\ 
     Qwen3-32B & \ding{55} & 70.9&	54.3&	17.8  \\ 
      \multicolumn{1}{c}{$\Delta$}  & &  {\color{green!75!black}+0.6}& {\color{red}-0.2}& {\color{green!75!black}+0.4} \\ \midrule
    \method-7B & \ding{51} & 72.0&	54.6&	17.7  \\ 
    \method-7B & \ding{55} & 68.2& 52.9		&16.4  \\ 
    \multicolumn{1}{c}{$\Delta$}   & & {\color{green!75!black}+3.9} & {\color{green!75!black}+1.7}  &{\color{green!75!black}+1.3} \\ \midrule

    Qwen2.5-7B & - & 65.5 & 50.1 & 15.0  \\ 
    \bottomrule
  \end{tabular}
  }
  \caption{\textbf{Performance with Reasoning activated and disactivated.} $\Delta$ is improvement with reasoning.}
  \label{tab:reason}
    \vspace{-3mm}
\end{table}

\paragraph{Reasoning Pattern} As illustrated in \cref{fig:reason_patter}, RL training profoundly impacts reasoning patterns. The trained model learns a strategic reranking approach, reasoning about the relevance of the passage to the query and extracting key judgment words. It also intelligently leverages terms like "\texttt{same}" for concise comparisons with prior passages, reducing verbose reasoning while still providing keywords. The model without RL training, even with a reasoning prompt, does not show such strategy and employs shorter reasoning chains.

\paragraph{Is improved reasoning transferable?} To investigate the transferability of improved reasoning gained from our reasoning-based reranking RL training, we evaluate the model on mathematical reasoning questions. Following \citep{ye2025limo}, we report the pass@1 score averaged over 16 samples with a temperature of 0.7 on the AIME 2024 and AMC datasets. As shown in Table~\ref{tab:math_reasoning_transfer}, we observe consistent improvements on both the challenging math tasks. This improvement suggests that training on the reranking task can, to a certain extent, transfer to other reasoning tasks.

\begin{table}[thb] 
\centering 
\resizebox{0.7\columnwidth}{!}{%
\begin{tabular}{lcc}
\toprule
Model & AIME 2024 & AMC \\
\midrule
Qwen2.5-7B-Instruct & 11.87 & 51.41 \\
(Ours) Rearank-7B & 12.92 & 52.66 \\
\bottomrule
\end{tabular}%
} 
\caption{Math Reasoning Transfer Results (Pass@1)}
\label{tab:math_reasoning_transfer}
   \vspace{-3mm}
\end{table}

\begin{figure}[!thb]
    \centering
    \includegraphics[width=\columnwidth]{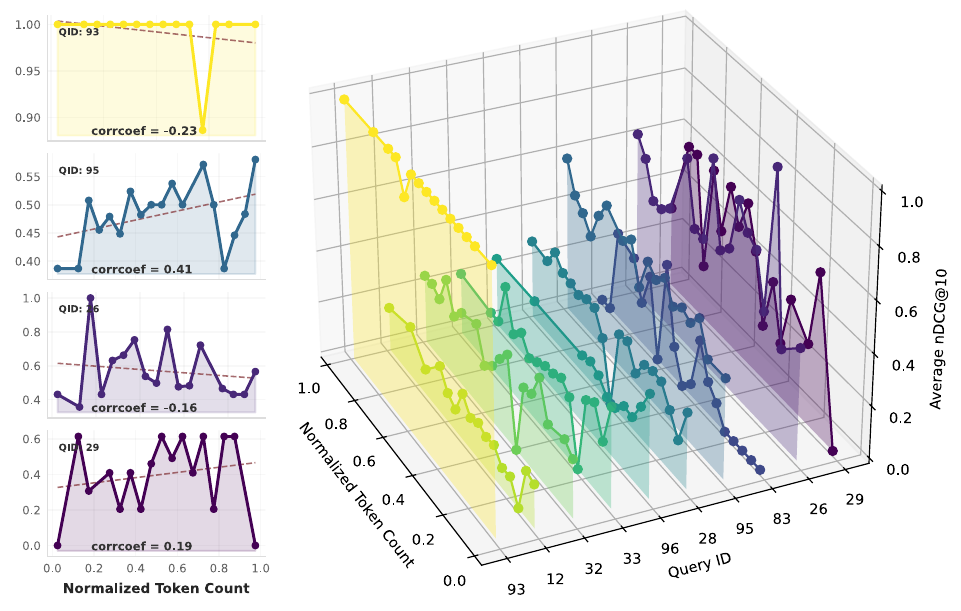}
    \caption{Impact of reasoning length. Data points are binned into 20 equal-width intervals by token count.}
    \label{fig:tokenlength}
    \vspace{-3mm}
\end{figure}

\paragraph{Impact of Reasoning Length on Performance}
We analyzed the impact of reasoning length on \method's reranking performance by repeatedly reranking biology samples from BRIGHT (50 times per query, temp. = 0.6), filtering instances with 0 scores. Contrary to \cite{marjanovic2025deepseek}, Figure \ref{fig:tokenlength} reveals no clear correlation between reasoning length and reranking performance.

\section{Conclusion}
We presented \method, a pioneering reasoning listwise reranking agent trained via Reinforcement Learning. \method~ significantly outperforms baselines and achieves competitive, even superior, results compared to GPT-4 and Qwen3 models across various benchmarks, notably with only 179 labeled training samples. Our analysis confirms that the RL-acquired reasoning capabilities transfer, facilitating new and effective reranking strategies.

\section*{Limitations}
While \method demonstrates promising results, several limitations should be acknowledged. Firstly, the quality and faithfulness of its generated explanations for ranking decisions, which may contain a certain degree of hallucination, have not been formally evaluated. Secondly, its performance heavily relies on the quality of initial candidates provided by BM25, which potentially limits improvements in scenarios with poor initial retrieval.

    \section*{Acknowledgement}
We sincerely appreciate the valuable feedback provided by Rabiul Awal, and Kanishk Jain, as well as the thoughtful input from all MAIR Lab members on multiple occasions. We thank the Mila IDT team and their technical support for maintaining the Mila compute cluster. This research was enabled in part by support provided by Calcul Québec and the Digital Research Alliance of Canada. We also acknowledge the material support of NVIDIA in the form of computational resources. Throughout this project, Aishwarya Agrawal received support from the Canada CIFAR AI Chair award.

\bibliography{custom}

\clearpage
\appendix

\section{Prompt Template}
\label{sec:prompt}
We provide the prompt template used in the experiments. Each prompt is concatenation of system prompt and user instruction. The user iteratively provides each passage paired with its original rank identifier. The prompt concludes with a post promt message containing the ranking query and explicit output format requirements.
\subsection{Reasoning prompt}

This is the prompt used by \method.
\begin{tcolorbox}[title=\textbf{Reasoning System Prompt}]
\small 
You are DeepRerank, an intelligent assistant that can rank passages based on their relevancy to the search query. You first thinks about the reasoning process in the mind and then provides the user with the answer.

I will provide you with passages, each indicated by number identifier []. Rank the passages based on their relevance to the search query.
Search Query: [QUERY].
Rank the [NUM] passages above based on their relevance to the search query.
The passages should be listed in descending order using identifiers. The most relevant passages should be listed first. The output format should be \texttt{<answer> [] > [] </answer>}, e.g., \texttt{<answer> [1] > [2] </answer>}.
\end{tcolorbox}

\begin{tcolorbox}[title=\textbf{Reasoning User Instruction}]
\small 

\textbf{Iterative User Message (per passage):}
\texttt{[RANK]} [Passage Content (truncated)]

\vspace{2mm} 

\textbf{Iterative Assistant Message (per passage):}
Received passage \texttt{[RANK]}.

\vspace{2mm} 

\textbf{Post Prompt:}
Please rank these passages according to their relevance to the search query: "[QUERY]"
Follow these steps exactly:
\begin{enumerate}
    \item First, within \texttt{<think>} tags, analyze EACH passage individually:
    \begin{itemize}
        \item Evaluate how well it addresses the query
        \item Note specific relevant information
    \end{itemize}
    \item Then, within \texttt{<answer>} tags, provide ONLY the final ranking in descending order of relevance using the format: \texttt{[X] > [Y] > [Z]}
\end{enumerate}

\end{tcolorbox}

\subsection{Zero-shot prompt}
This is original prompt used for RankQwen and RankGPT.
\begin{tcolorbox}[title=\textbf{Zero-shot System Prompt}]
\small 
You are RankGPT, an intelligent assistant that can rank passages based on their relevancy to the query. I will provide you with [NUM] passages, each indicated by number identifier []. \newline Rank the passages based on their relevance to query: [QUERY].
\end{tcolorbox}

\begin{tcolorbox}[title=\textbf{Zero-shot User Instruction}]

\small 

\textbf{Iterative User Message (per passage):}

\texttt{[RANK]} [Passage Content (truncated)]

\vspace{2mm} 

\textbf{Iterative Assistant Message (per passage):}
Received passage \texttt{[RANK]}.

\vspace{2mm} 

\textbf{Post Prompt:} Search Query: [QUERY]. \newline Rank the [NUM] passages above based on their relevance to the search query. The passages should be listed in descending order using identifiers. The most relevant passages should be listed first. The output format should be \texttt{[] > []}, e.g., \texttt{[1] > [2]}. Only response the ranking results, do not say any word or explain.
\end{tcolorbox}

\begin{figure}[htb]
    \centering
    \includegraphics[width=\columnwidth]{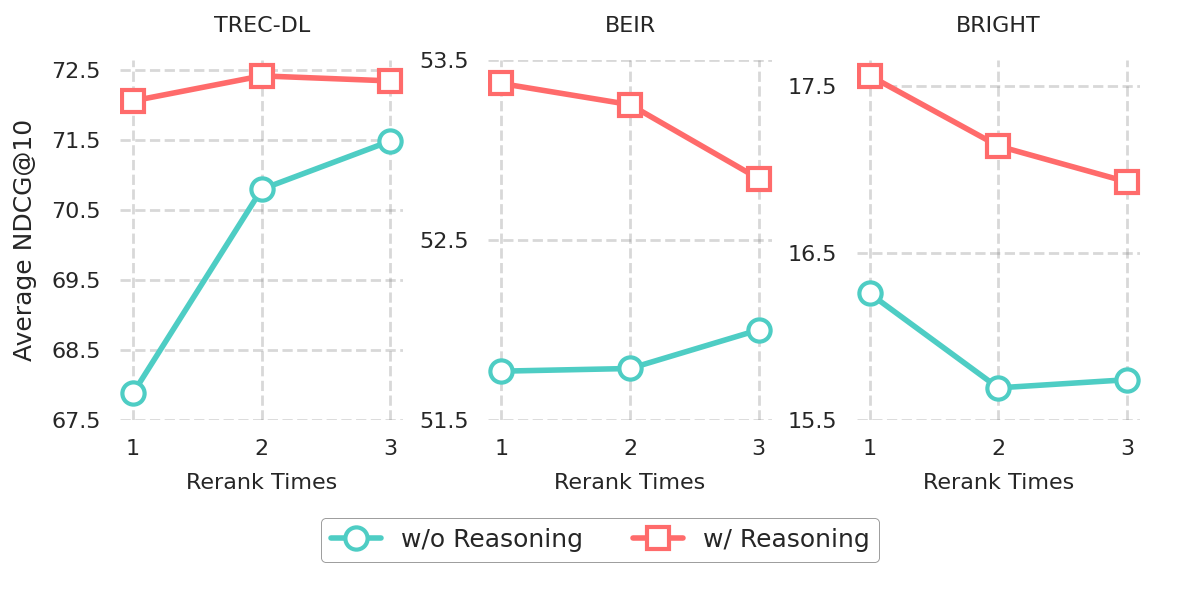}
    \caption{Multiple reranking results of \method.}
    \label{fig:multirerank}
    \vspace{-5mm}
\end{figure}

\paragraph{Multiple Rerank Pass}

Analyzing the impact of multiple reranking passes with \method (evaluating performance with and without reasoning) reveals mixed results across benchmarks (\cref{fig:multirerank}). Multiple passes improve performance on TERC-DL and BEIR without reasoning, but degrade it on BRIGHT. With reasoning, gains are seen on TREL-DL, but performance is harmed on other benchmarks. These findings indicate that for \method, a single reranking pass is sufficient.

\section{Case Study}
\begin{figure*}[thb]
    \centering
    \includegraphics[width=\linewidth]{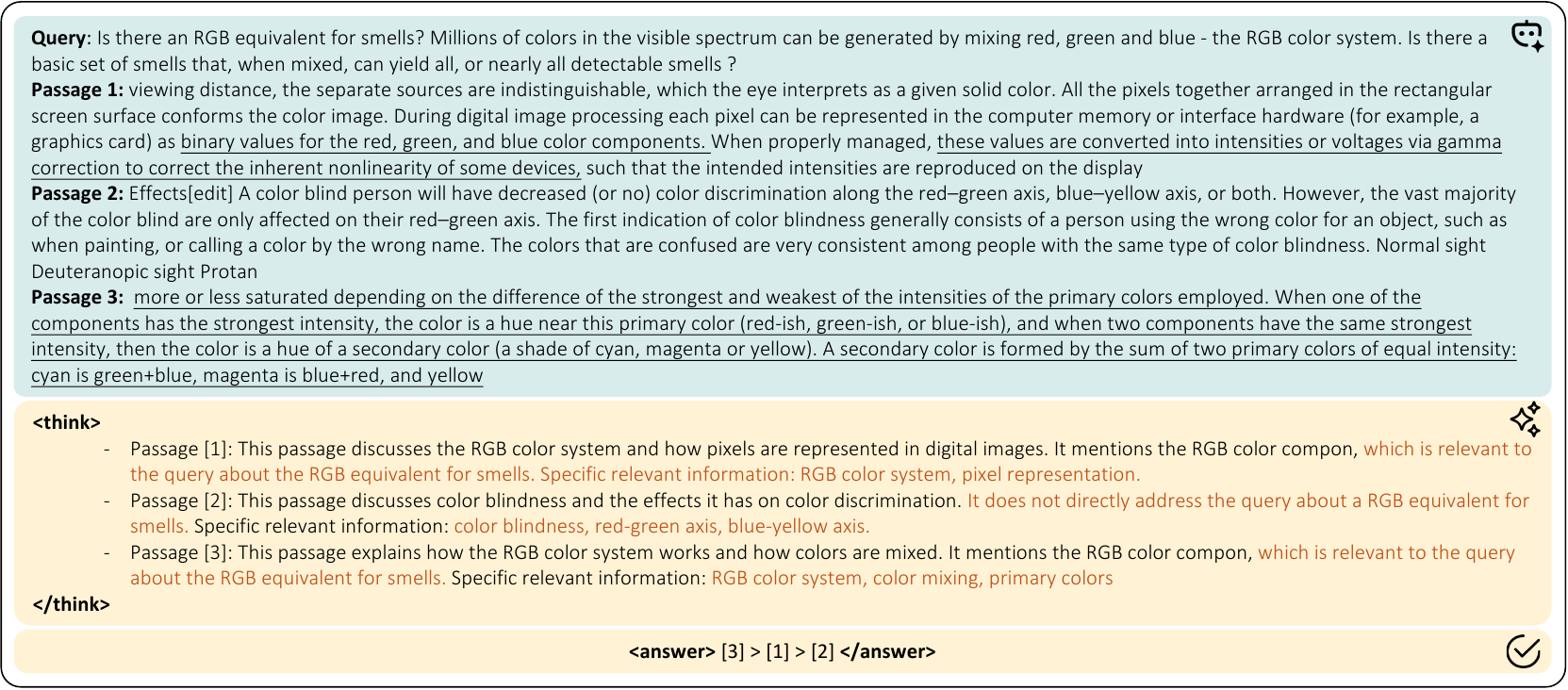}
\caption{Examples from biology split of BRIGHT, only 3 passages selected for better visibility. Relevant sentences in the passages and Key reasoning content in the thinking process are highlight. }
    \label{fig:example}
    \vspace{-5mm}
\end{figure*}

Figure \ref{fig:example} provides an illustrative example from the BRIGHT dataset, detailing the inference reasoning process employed by \method. For each passage, the agent first analyzes its content to understand its key themes and then determines its relationship to the query. Relevant information, typically keywords, is then extracted based on this analysis. We conducted a manual evaluation of the generated reasoning content and found it to be of high quality, contributing to the trustworthiness of the system's outputs. This reasoning process ultimately informs the final ranking of passages presented in the answer.

\end{document}